# Silicon Photomultipliers in the CMS Upgrade

**A. Heering[a], A. Karneyeu[a], Y. Musienko[a,b,1] and M. Wayne[a,2]**

[a] *University of Notre Dame,*
*Notre Dame, IN, U.S.*
[b] *Institute of Nuclear Research,*
*Moscow, Russia*
*E-mail*: mwayne@nd.edu, iouri.musienko@cern.ch

ABSTRACT: Silicon Photomultipliers (SiPM) will be used extensively in the upgraded CMS detector at the Large Hadron Collider. SiPMs have already been implemented into the barrel and endcap hadron calorimeters (HCAL) as part of the Phase I upgrade, and hundreds of thousands of SiPMs will be used for two new Phase II subdetectors, the Barrel Timing Layer (BTL) and the endcap high granularity hadronic calorimeter (HGCAL). We discuss the motivation for SiPMs as the photodetectors of choice in CMS, the evolution of the SiPMs from Phase I to Phase II, and the particular challenges faced and overcome for each of the three subdetectors.

---

[1] Corresponding author.
[2] Corresponding author.

## Contents

## 1. Introduction

Silicon Photomultipliers (SiPMs) are pixelated APDs that operate in Geiger mode. Each cell of the SiPM is separated from the bias voltage by a current limiting quench resistor. The absorption of a photon of proper wavelength can excite an avalanche that causes the struck cell to discharge. The firing of a cell causes its capacitor to discharge, creating a quantized charge output from the SiPM depending on the number of cells that discharge. The SiPMs have low operating voltage, high quantum efficiency and gain (both values are superior to that of HPDs), low excess noise factor even at high gain (excess noise factor of APDs increases with the gain), and insensitivity to magnetic fields [1, 2]. Over the past decade, the CMS collaboration decided to implement more than 500,000 channels of SiPMs as part of the Phase I (during two long-term shutdowns of the LHC in 2013–2015 and in 2019–2022) and Phase II upgrades of the detector (planned during the LHC shutdown in 2026-2030). SiPMs are the photodetectors of choice for three important subsystems in CMS: the hadronic calorimeter, made up of outer, barrel and endcap subdetectors; a new endcap detector that integrates electromagnetic and hadronic calorimetry together; and a Barrel Timing Layer that is part of a new MIP Timing Detector for high luminosity running at the LHC [3, 4]. In this paper we discuss the motivation for SiPMs as the photodetectors of choice in CMS, the evolution of the SiPMs from the Phase I to Phase II upgrades, and the particular requirements placed on the SiPM performance for each of the three subdetectors. While the basic photodetector technology chosen is the same for all three subsystems, each presented their own set of challenges (to be discussed in the following chapters).

## 2. Phase I Upgrade of the Hadronic Calorimeter

The current hadron calorimetry (HCAL) for CMS is made up of large scintillating tiles read out by Y11 wavelength shifting fibers [5]. At the time of the detector design, circa 2000, the best choice for the HCAL photodetectors were Hybrid Photodiodes (HPD). These devices have reasonable photodetection efficiency (PDE) of about 12%, are fast and have a large dynamic range. Their relatively large size limited the channel count, and consequently the depth segmentation for the barrel and endcap detectors. The main goals of the HB and HE upgrades were to replace the HPD photodetectors, which produced anomalous signals [6] and showed signal degradation [7], and to increase the HCAL depth segmentation.

With the development of SiPMs it became clear that the HCAL performance would be greatly improved by replacing HPDs with this new technology. The SiPMs have low operating voltage, high quantum efficiency and gain (both values are superior to that of HPDs), low excess noise factor even at high gain (excess noise factor of APDs increases with the gain), and insensitivity to magnetic fields. Additionally, their compact size enables improved depth segmentation and they are insensitive to the 3.8 Tesla magnetic field at CMS [5]. The most important challenges of the SiPM operation in the CMS HCAL were: large dynamic range of signals produced by hadrons in the calorimeter (from tens to tens of thousand of photoelectrons) and their high rate (due to 40 MHz LHC proton collision rate). The SiPMs were required to operate at room temperature and remain operational at high neutron fluences (<10 % change of gain and PDE, moderate noise increase after 1 x $10^{12}$ neutrons/cm$^2$, 1 MeV equivalent and 14 Gray ionizing dose). To reach such a performance the SiPMs need to have small cell size and fast cell recovery time. At the same time the PDE of SiPM should be kept high (>20%) which is difficult to achieve due to increased non-sensitive area in the case of small cells.

The SiPM replacement in HCAL was done in three steps. First, about 2,000 SiPMs were installed into the outer barrel (HO). These devices were developed and fabricated at Hamamatsu (S10931-50, a smaller area SiPM of this type had been already used in T2K experiment [8]). They have 3x3 mm$^2$ active area and 50 μm cell size, PDE=30% (for Y11 light, peak wavelength ~500 nm)) and gain=0.9 x $10^6$ at 3 volts overvoltage. At the time there were no SiPMs available with small cell size and high PDE. The HO is a “tail catcher” – part of the CMS HCAL that is out of solenoid. The range of signals in HO is small and the detector does not require small cell SiPMs. For the Endcap (HE) and Barrel (HB) calorimeter upgrades we again used Hamamatsu SiPMs (S12571). At the time, these SiPMs had the highest PDE and smallest dark noise increase after irradiation among existing types of 15 μm cell size SiPMs from all the producers [5, 9, 10]. The SiPMs have circular active areas with diameters of either 2.8 or 3.3 mm and 15 μm cell size for improved dynamic range [4] These SiPMs have PDE=32% and gain=0.3 x $10^6$ at 3 volts over breakdown. The total number of SiPMs used in the combined upgrades of HE and HB is about 25,000. Figure 1 shows several HO SiPMs mounted on their PC board on the left, and HE and HB SiPM arrays in the 8-channel package designed by the Notre Dame group on the right.

The quality control program was identical for the HE and HB upgrades. We measured the breakdown voltage, signal response (SiPM current under stable light illumination at 3V over the breakdown voltage), dark current, forward resistance (together with cell capacitance, the forward resistance divided by number of cells determines the cell recovery time), and capacitance for each channel. A total of 1400 production HE arrays were tested, and 104 arrays, or 7.43%, were rejected. Since there are eight SiPM channels per array, this corresponded to a yield of good SiPMs better than 99%. For the HB production, a total of 1680 production arrays went through the same quality control testing, with only 82 rejected, again giving a yield of good SiPMs better than 99%. The uniformity was also excellent throughout the two production runs. Figure 2 shows the distributions of breakdown voltage and the spread in relative response for uniform LED illumination for a sample of 500 arrays (4000 channels) measured at 3V over breakdown. The 515 nm wavelength LED’s intensity was chosen high enough to accurately determine the breakdown voltage of the SiPMs. The SiPMs current was about 10 μA at an overvoltage of 3 volts.

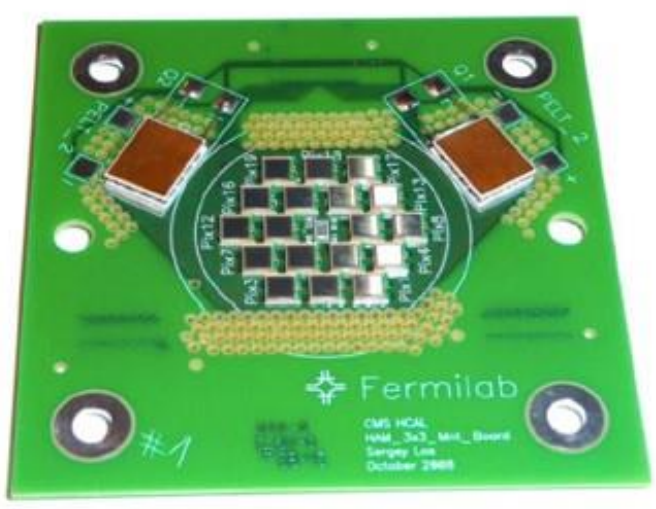

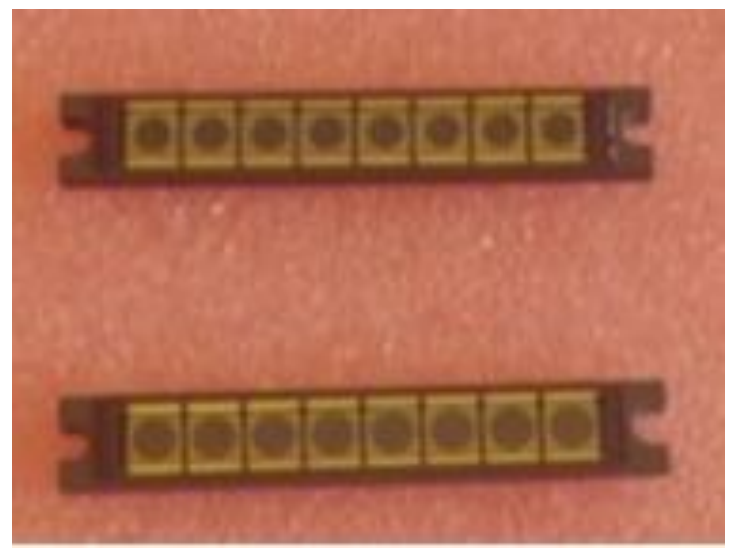

**Figure 1.** Left: an HO 18-channel SiPM board showing several SiPMs in the central region of the board. Right: HB and HE eight-channel arrays with the individual SiPMs mounted on a ceramic package.

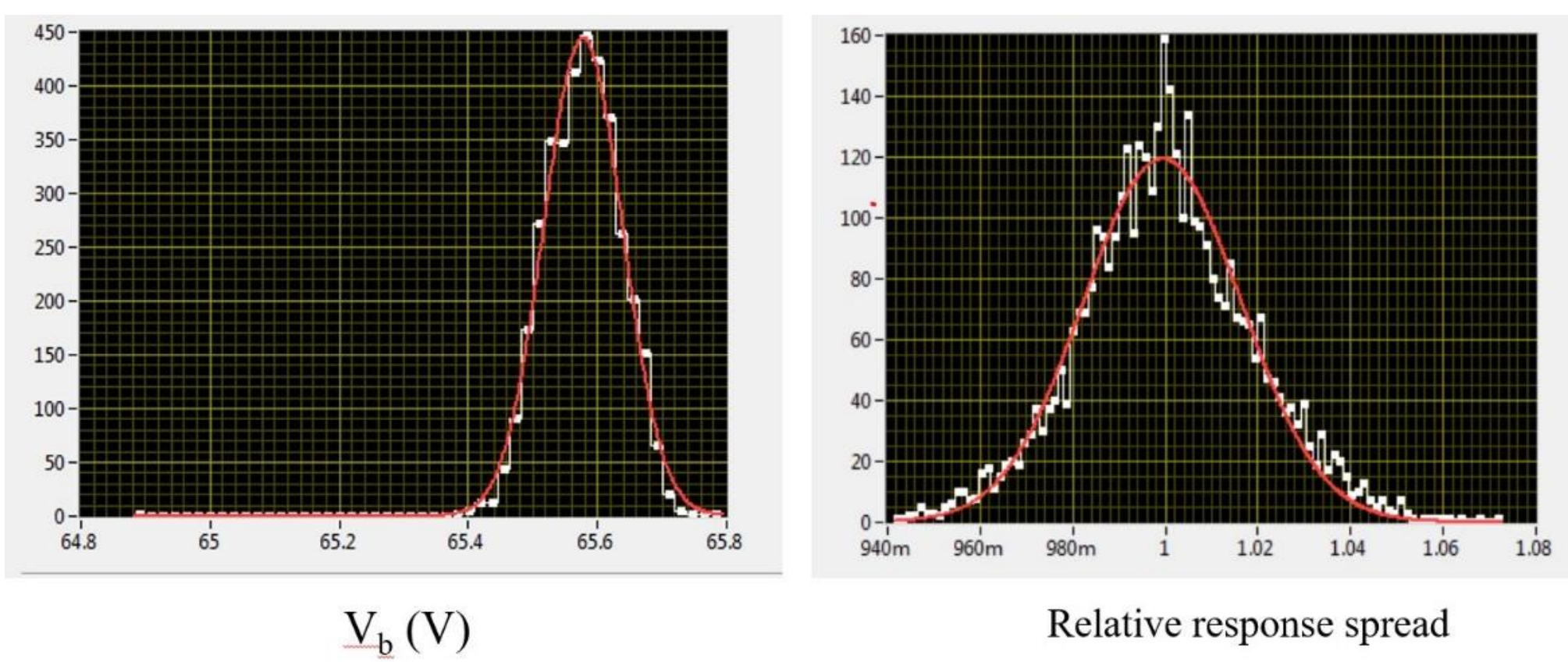

**Figure 2.** Quality control results for 500 arrays (4,000 channels) of HE SiPM. Left: the distribution of breakdown voltage ($V_b$). Right: the spread in relative response for uniform LED illumination.

## 3. Phase II Upgrades

### 3.1 CMS Endcap Calorimeter

As part of the ongoing CMS Phase II upgrade project, the endcap electromagnetic and hadronic calorimeters will be replaced by a single new detector, the high granularity endcap calorimeter or HGCAL[11]. The electromagnetic section of HGCAL has layers of silicon hexagonal structures as the active element, sandwiched between absorbers. The hadronic section also uses silicon as the active material at low eta, and layers of scintillator tiles read out by SiPMs at higher pseudorapidity [11]. The scintillator part of the detector contains about 240,000 individual tiles ranging from 4 to 30 cm$^2$ in area, each coupled to a SiPM with 9 mm$^2$ active area and 15 µm cell size. Figure 3 shows a schematic of the HGCAL along with photos of a tileboard, individual scintillator tiles and the SiPM package.

Requirements for the scintillator part of the HGCAL include:

- High granularity for particle flow reconstruction and pileup mitigation.
- Maintaining the ability to calibrate individual tiles using minimum-ionizing particles through the life of the calorimeter for the full HL-LHC operation, with expected fluence up to 5E13 1-MeV neutrons/cm$^2$ equivalent.
- The HGCAL must fit within the existing envelope of the current detectors.

The R&D on HGCAL SiPMs built upon the previous work done for the Phase I HCAL upgrade, with particular emphasis on radiation hardness since the expected end-of-life doses in HGCAL will be 50 times higher (the most important challenge we faced). We worked with multiple vendors and after a market survey chose Hamamatsu and their S16713-3 SiPM for their high PDE, fast cell recovery time, and good resistance to radiation. SiPM characteristics included: 3x3 $mm^2$ sensitive area, 15 um cell size, PDE(420 nm) = 25 %, cell recovery time ~10 ns, gain = 2.3 x $10^5$, and signal response reduction < 15% after 5E13 n/$cm^2$. All parameters are given for 2 Volts overvoltage at T=-35 °C. Production started in summer 2024 and by the end of the year we had received about 50% of the devices, with the rest expected by spring 2025.

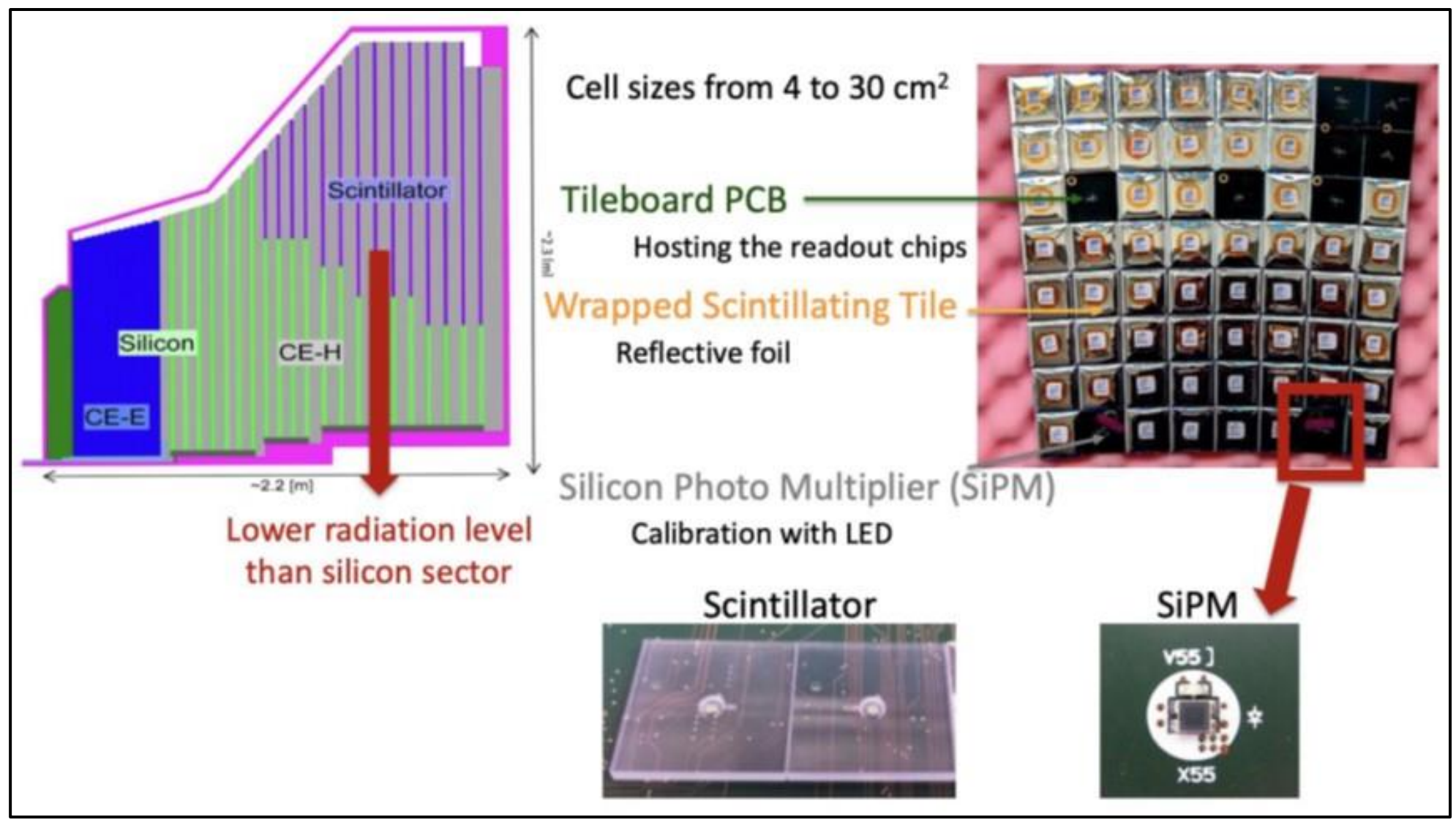

**Figure 3.** Upper left: Schematic of the HGCAL detector. Upper right: tileboard PCB with wrapped tiles. Lower right; individual SiPM on package. Lower center; unwrapped scintillator tiles.

Each of the more than 300,000 SiPMs is mounted in a SMD package package (designed by the producer), arriving on reels of 1,000 channels. Unlike the HCAL upgrade, where we tested every channel, we are only taking a 5% sample for quality control and characterization, with the other 95% going directly to detector assembly. For this sample we measure IV curves with and without LED illumination to measure dark current, breakdown voltage and signal response. We also measure forward resistance to verify the quenching resistance. The results are compared with a set of specifications developed together with Hamamatsu, and the yield of good SiPMS has been well over 99%. Measurements of PDE, gain and capacitance are performed on a subset of channels, and approximately 1% of the total is set aside for irradiation and other long-term tests. For radiation testing SiPMs are exposed to neutrons at the JSI reactor in Ljubljana, annealed to reduce the dark current and remeasured at about -30ºC, the expected operating temperature for HGCAL.

Good performance after irradiation is critical for the HGCAL SiPMs due to the high radiation levels expected in the detector, as discussed above. Figure 4 shows results before and after exposure to 5E13 1-MeV neutrons/$cm^2$ equivalent at JSI. The results for PDE, gain and signal amplitude are all within 10-15% of their unirradiated values in the expected operating range of 1.0 - 1.5 volts over breakdown. The fourth plot shows that the signal-to-noise ratio is essentially flat over a relatively wide range of operating voltages around 1.0 volts. For these tests the SiPMs were irradiated in groups of 8 channels and the uniformity of signal response after radiation was within a few percent.

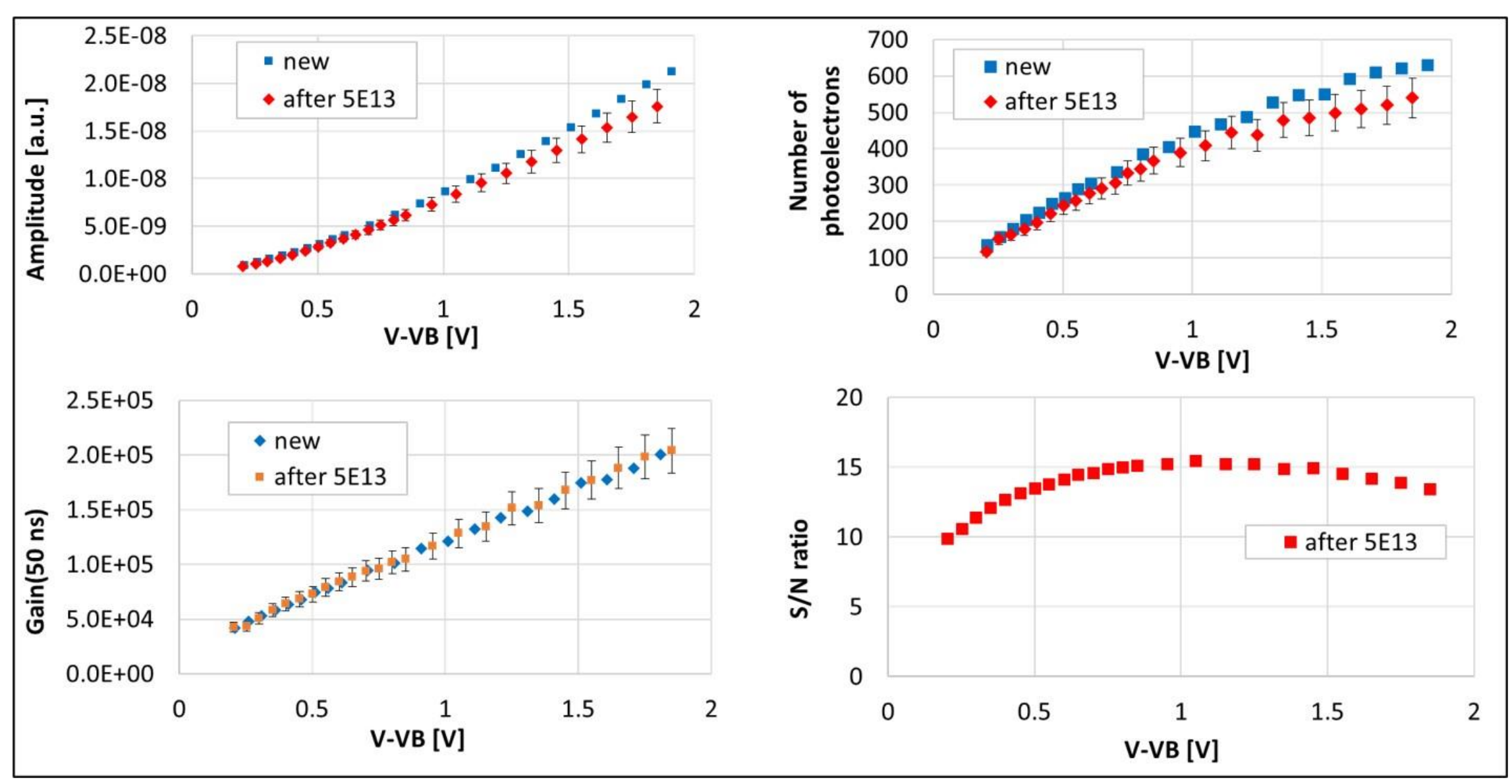

**Figure 4.** Comparisons of HGCAL SiPM performance before and after irradiation to 5E13. Upper left: signal amplitude. Upper right: number of photoelectrons detected. Lower left: gain (measured within a 50 ns gate). Lower right: signal-to-noise ratio as a function of voltage above breakdown. The SiPM was illuminated with 420 nm LED pulsed light (~2500 photons/pulse, 50 ns signal integration time).The measurements were taken at a temperature of -36$^{o}$C.

### 3.2 CMS Barrel Timing Layer

Part of the CMS Phase II upgrade will be the addition of a MIP Timing Detector (MTD) [12]. By dividing the interaction region into time slices, the ability to identify and separate vertices will be improved during the high pileup environment of the HL LHC. Simulations show that this will benefit many future physics analyses. The new detector will have both a Barrel Timing Layer (BTL) and Endcap Timing Layer (ETL). The BTL will contain a single layer of approximately 165,000 LYSO crystal bars with dimensions of 55.0 x 3.1 x 3.8 mm$^3$. Each crystal is read out with a SiPM at both ends. The LYSO crystals are assembled into layers of 16 parallel bars, and the SiPMs attached to each end are packaged into 16-channel arrays (Fig.5). Our group designed the 16-channel package, including flex circuit, that will house the ~330,000 SiPMs.

The BTL presents unique challenges for choice of photodetector. The expected radiation fluence at the end-of-life is 2E14 1-MeV neutrons/cm$^2$ equivalent, so four times the level expected for HGCAL and 200 times that of the HB detector. Also, since the BTL is a timing detector we need fast signals with relatively large amplitudes. To combat the severe radiation environment, we came up with a plan to add four small thermoelectric coolers (TECs) to each package, as seen in Figure 5. The TECs enable an additional temperature reduction of 10$^{o}$C during BTL operation, down to -45$^{o}$C. This extra cooling is "power neutral" as the additional power needed for the TECs is offset by the reduction in dark current of the SiPMs. By reversing the bias voltage to the TECs during detector downtimes, we will use them to heat the SiPMs (up to 60$^{o}$ C) and significantly anneal radiation damage (dark current will be reduced by a factor of 6). Using both cooling and annealing we estimate the TECs will reduce the SiPM dark current by a factor of 10-12 throughout the lifetime of the experiment.

Even with the significant benefit provided by the TECs, initial studies done with SiPMs of 15 µm cell size were not able to demonstrate the required timing resolution for the BTL. We designed a simulation of the SiPM response, folded that into the downstream electronics response and found that increasing the cell size would result in improved timing for the BTL. Working with our SiPM vendors (Hamamatsu and FBK) we measured the timing, in a beam test, for a range of SiPM cell sizes, from 15 to 30 µm in steps of 5 µm [13]. As a result, 25 µm was chosen as the optimal cell size because of the best timing resolution achieved among irradiated (2E14 n/cm$^2$) SiPMs.

Working with CERN, we carried out a market survey of SiPM vendors and developed the final set of specifications for the SiPM performance. The most important of them are: PDE(420 nm) > 48 %, gain > 7x10$^5$, cell recovery time < 10 ns (all parameters are given at an overvoltage of 3 volts). After an extensive set of measurements and comparisons, including thorough radiation studies of the candidate SiPMs [14], we selected the S15408-4125TC SiPM from Hamamatsu for the BTL. The SiPMs have 25 um cell size and an active area of 2.9 mm x 3.8 mm to match the dimensions of the LYSO crystals. The BTL SiPM fabrication began in early 2024, and by the end of the year more than 50% of the total production had been delivered to CERN. The delivery of the final batch of SiPMs is expected in spring 2025.

The quality control of these SiPMs is shared equally between our group at CERN and colleagues at Debrecen University in Hungary. For the BTL, every channel that goes into the detector must be measured and characterized. To accommodate the 16-channel array design and ensure that we can test the large total number of channels on schedule, our group designed a new test system capable of testing 12 arrays at once, quickly switching through the channels as we make the required measurements. Measurements include IV curves, forward resistance, capacitance and determination of the breakdown voltage. Because all 16 SiPMs in an array share a single bias voltage, the spread in breakdown voltage in the array is required to be ≤ 150 mV. This specification has been met for each of more than 10,000 arrays tested so far, and the overall per-channel yield for all specifications is above 99%. As with the HGCAL project, a small subset of channels is set aside for radiation testing and accelerated aging studies

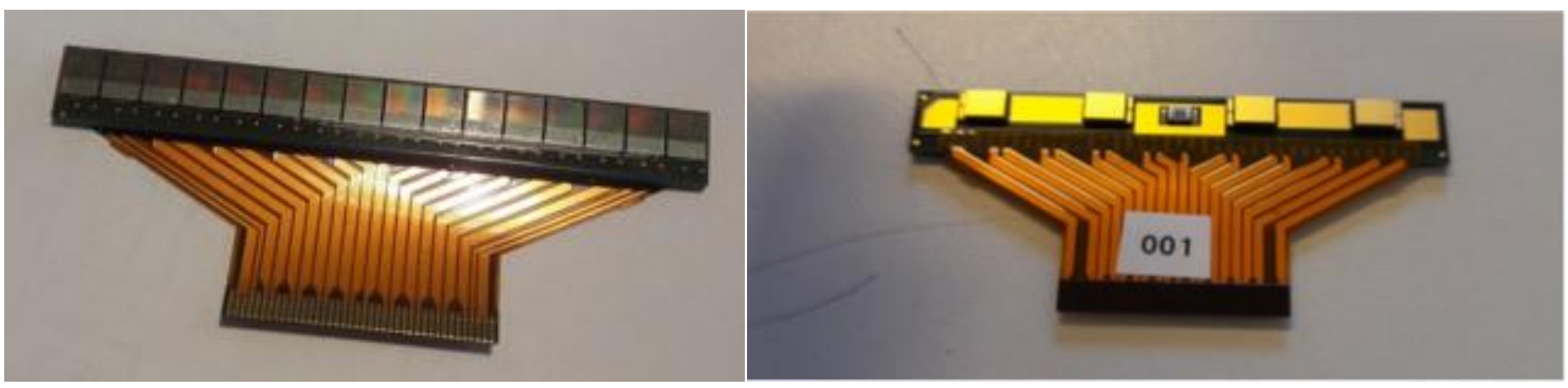

**Figure 5.** The BTL 16-channel SiPM array. Left: the top side of the array with the 16 individual SiPMs visible. Right: the back side of the array showing the four TECs and an RTD for temperature monitoring.

## 4. Summary

Starting more than a decade ago, our group has worked with vendors to help develop, package, fabricate and implement hundreds of thousands of Silicon Photomultipliers (SiPM) into three major components of the CMS detector upgrade at CERN. We demonstrated the benefits of SiPM technology in the Phase I upgrade of the hadronic calorimeter, and building off that success CMS is implementing SiPMs into two new detectors as part of the Phase II upgrade. Each of the three projects presented their own unique set of challenges for the SiPMs, based upon the performance requirements of each detector and its radiation environment. An extensive program of quality assurance and control carried out at CERN has shown that large quantities of SiPM can be fabricated with very high yields (> 99%) and performance within the tight range of specifications required by the physics needs of the CMS experiment.

## Acknowledgments

We would like to acknowledge our CMS colleagues on the HCAL, HGCAL and MTD upgrade projects. This project has received funding from the European Union's Horizon Europe Research and Innovation programme under Grant Agreement No 101057511 (EURO-LABS). We also thank the Department of Energy and the National Science Foundation for their support of this work.